\documentclass[a4paper, 12pt]{article}

\usepackage{graphicx}%
\usepackage{multirow}%
\usepackage{amsmath,amssymb,amsfonts}%
\usepackage{amsthm}%
\usepackage{mathrsfs}%
\usepackage[title]{appendix}%
\usepackage{xcolor}%
\usepackage{textcomp}%
\usepackage{manyfoot}%
\usepackage{booktabs}%
\usepackage{algorithm}%
\usepackage{algorithmicx}%
\usepackage{algpseudocode}%
\usepackage{listings}%
\usepackage{physics}
\usepackage{times}
\usepackage{authblk}
\usepackage[nodisplayskipstretch]{setspace}
\usepackage{geometry}
\usepackage{natbib}
\usepackage{xifthen}
\usepackage{pgffor}

\newcommand{\citetbjps}[2][]{\ifthenelse{\equal{#1}{}}{\citeauthor{#2} ([\citeyear{#2}])}{\citeauthor{#2} ([\citeyear{#2}], #1)}} %Jones ([2000]) or Jones ([2000], p.41)
\newcommand{\citepbjps}[2][]{\ifthenelse{\equal{#1}{}}{(\citeauthor{#2} [\citeyear{#2}])}{(\citeauthor{#2} [\citeyear{#2}], #1)}} %(Jones [2000]) or (Jones [2000], p.41)

\newcommand{\citealtbjps}[2][]{\ifthenelse{\equal{#1}{}}{\citeauthor{#2} [\citeyear{#2}]}{\citeauthor{#2} [\citeyear{#2}], #1}} %Jones [2000] or Jones [2000], p.41

\usepackage{float}
\usepackage{wrapfig}
\usepackage{caption}
\usepackage{enumitem}
\setlist[itemize]{leftmargin=*}
\setlist[enumerate]{leftmargin=*}
\usepackage{pdfcolfoot}
\usepackage[explicit]{titlesec}
\titlespacing*{\section}
{0pt}{0.33cm}{0.33cm}
\titlespacing*{\subsection}
{0pt}{0.33cm}{0.33cm}

\usepackage{url}
\titleformat\section[block]%
  {\bfseries\centering}
  {\thesection}
  {\baselineskip}
  {{#1}}
\usepackage{ragged2e}
\usepackage[hang]{footmisc}
\usepackage{csquotes}
\renewenvironment{quote}{\list{}{\rightmargin 1.1em\leftmargin1.1em}\item\relax
\small\singlespacing}{\endlist}
\SetBlockEnvironment{quote}

\begin{document}

%\onehalfspacing
%\doublespacing
\title{Temporal Nonlocality from Indefinite Causal Orders}
\author[1]{Laurie Letertre}
\affil[1]{\small Munich Center for Mathematical Philosophy, Ludwig-Maximilians-Universität München, Geschwister-Scholl-Platz 1, München D-80539, Germany}
%laurie.letertre@gmail.com
\date{Forthcoming in \textit{The British Journal for the Philosophy of Science}}

%\affil[2]{\orgdiv{Institute of Philosophy}, \orgname{Czech Academy of Science}, \orgaddress{\street{Jilská 1}, \city{Prague}, \postcode{11000}, \state{}, \country{Czech Republic}}}
\maketitle

\begin{abstract}
A temporal counterpart to Bell nonlocality would intuitively refer to the presence of non-classical correlations between timelike-separated events. The hypothesis of temporal nonlocality has received recent support in the literature, and its existence would likely influence the future development of physical theories. This paper shows how Adlam's principle of temporal locality can be violated within a protocol involving indefinite causal orders. While the derivations of Leggett-Garg inequalities or the temporal CHSH inequality are said to involve problematic assumptions preventing a targeted probing of a well-defined notion of temporal nonlocality, the present test is free from such worries. However, it is shown that the test, in its current formulation, fails to be fully model-independent. We provide several considerations regarding the physicality of ICOs that could help alleviate this drawback. In the present work, a specific physical interpretation of ICOs in terms of retrocausal influences would explain the presence of temporally nonlocal correlations. It is argued that, as the physical underpinnings of temporal nonlocality might also account for standard Bell nonlocality, focusing on the former as a consequence of ICOs might support under-explored strategies to make sense of the latter.
\end{abstract}

\vspace{0.5cm}
\textbf{Keywords: }Temporal nonlocality, indefinite causal order, classical order, quantum nonlocality
\newpage
\section{Introduction}\label{intro}

Bell nonlocality can be understood as taking place across space, as it refers to the presence of non-classical correlations across spacelike-separated events~(\citealtbjps{bell1964einstein}; \citealtbjps{bell1976}). These correlations violate the so-called Bell inequalities, which are derived upon assuming, among other auxiliary premises, the principle of local causation: `If two sets of space-like physical events A and B are correlated, then there exists a set of events C in the past common to A and B such that conditioning on this set eliminates the correlation'~(wording taken from \citepbjps[p.~13]{wiseman2017causarum}). The existence of Bell nonlocality has been experimentally demonstrated~\citepbjps{aspect1982experimental}. 
Its postulated origin, or underlying mechanism, will depend on the interpretation one gives to the theory of quantum mechanics (QM),\footnote{If QM is understood as referring to our relation to the external world (i.e. when a non-representationalist interpretation of QM is adopted), Bell nonlocality is understood as affecting the way we relate to the world. If the formal objects of QM are taken to refer to elements of the external world, nonlocality will be a true feature of objective reality.

In the latter case, some approaches will embrace the existence of Bell nonlocality \textit{per se} and provide some dynamical mechanisms underlying it (e.g. Bohmian mechanics~\citepbjps{bohm1952suggested}, or the GRW theory~\citepbjps{PhysRevD.34.470}), while others will rather reject a peripheral premise used to derive the Bell inequalities, such as the unicity of the measurement outcomes in an experiment (i.e. the many world interpretation of QM, see e.g. \citepbjps{wallace2012emergent}), or the free choice of its measurement settings (i.e. the superdeterminist approaches, see e.g. \citepbjps{baas2023does}).} leading to various understanding of the phenomenon. Importantly, the experimental test of Bell nonlocality is model-independent: it does not rely on the validity of any particular physical model. By contrast, quantum nonseparability (or entanglement) is a model-dependent concept, necessary but non-sufficient to observe Bell nonlocality. It is defined within the theoretical framework of QM, and describes correlations among quantum states of distinct systems at the same time.

In that context, previous works explored the possibility of a temporal counterpart to Bell nonlocality, one that would refer to the presence of non-classical correlations between timelike-separated events. The following questions emerge: 

\begin{itemize}
 \setlength\itemsep{0.05em}
    \item[-] How should we define temporal nonlocality?
    \item[-] Would this principle connect with a form of temporal entanglement ? 
    \item[-] Could we test for its existence? 
    \item[-] How would temporal nonlocality relate to spatial (i.e. Bell) nonlocality? 
\end{itemize}

There are various motivations to investigate the hypothesis of temporal nonlocality. On the one hand, there exists some support for its existence in nature. First of all, \citetbjps{adlam2018spooky} makes the case that Bell's theorem assumes both spatial and temporal locality. Violations of Bell inequalities give therefore as much evidence against spatial locality than against temporal locality. Temporal locality has been assumed rather uncritically for historical and pragmatic reasons rather than for good scientific ones. \textcolor{black}{Moreover, she points out that, according to the theory of relativity, an instance of spatial nonlocality becomes an instance of temporal nonlocality upon a change of reference frame. The assumptions of spatial nonlocality, temporal locality and relativity are therefore incompatible}. Another support for the hypothesis of temporal nonlocality comes from \citetbjps{leifer2017time}, who proposed an argument according to which time-symmetric quantum mechanics must entail retrocausality. \citetbjps{rodriguez2023temporal} modified this argument, and showed that time-symmetric QM more generally entails temporal nonlocality as defined in \citepbjps{adlam2018spooky}. \textcolor{black}{Finally, as discussed in more detail below, delayed-choice entanglement swapping experiments can be seen as providing support to some form of temporal nonlocality.}
On the other hand, as discussed by \citetbjps{adlam2018spooky}, the existence of temporal nonlocality would impact the way we make progress in physics, and should therefore be carefully examined. For example, temporal nonlocality can possibly support under-explored interpretations of quantum mechanics, e.g. involving retrocausal influences, non-Markovian dynamics or atemporal models. Furthermore, the question of temporal nonlocality is of foundational interest, as it would clarify how quantum phenomena relate to the spatial and temporal dimensions, which might impact future developments in more fundamental theories.

The ideal scenario to test for temporal nonlocality would be that, given (i) a specific experimental setup with classical inputs and outputs of measurements, (ii) a set of reasonable hypotheses, and (iii) a suitably defined principle of temporal locality, temporal inequalities would be derived. If these inequalities were violated by the probability distributions of measurements' outcomes, we could infer that the principle of temporal locality needs to be rejected, and we would speak of temporally nonlocal correlations. Different kinds of temporal inequalities have been proposed in the literature~(\citealtbjps{leggett1985quantum}; \citealtbjps{brukner2004quantum}). In \citepbjps{brukner2004quantum}, the relevant principle for temporal locality is called locality in time, and states that `the results of a measurement performed at time T2 are independent of any measurement performed at some earlier or later time T1'. A temporal version of the CHSH inequality is obtained. 
In~\citepbjps{leggett1985quantum}, inequalities are derived, not to probe the existence of some kind of temporal nonlocality, but instead to probe the quantumness of macroscopic systems. The core principle needed for the derivation is that of macrorealism, which states that “a macroscopic object, which has available to it two or more macroscopically distinct states, is at any given time in a definite one of those states, and it is possible in principle to determine which of these states the system is in without any effect on the state itself, or on the subsequent system dynamics”. Still, Leggett-Garg inequalities are often called temporal Bell inequalities, because the experimental setup used to test the violation of these inequalities involves a sequence of measurements of the same observable at different times.\footnote{There exist different derivations of the inequalities, see e.g.~(\citealtbjps{rastegin2014formulation}; \citealtbjps{clemente2016no}; \citealtbjps{ali2023bell}). Relatedly, \citetbjps{maroney2014quantum} argue that a careful analysis of Leggett-Garg inequalities shows that their violation can only imply a rejection of a weaker reformulation of macrorealism, stating that `the only possible preparation states of a system S are operational eigenstates
of [the quantity Q to be measured] and statistical mixtures thereof'. This result amounts to say that the inequalities test whether one can understand the indeterminacy of a measurement's result as classical ignorance.} Instead of the spacelike separated parties found in Bell’s theorem, we deal with timelike-separated parties. 
Numerous experiments have been made to test whether one can empirically observe a violation of the temporal version of the CHSH inequality~\citepbjps{ringbauer2018multi}, or of the Leggett-Garg inequalities~(\citealtbjps{PhysRevLett.107.130402}; \citealtbjps{goggin2011violation}; \citealtbjps{dressel2011experimental}; \citealtbjps{knee2012violation}; \citealtbjps{emary2013leggett}).\footnote{A discussion regarding the potential loopholes in these experiments is available in \citepbjps{vitagliano2023leggett}.}

There are reasons why the inequalities derived from 'locality in time' in \citepbjps{brukner2004quantum}, or from 'macrorealism' in \citepbjps{leggett1985quantum} might not be satisfying tests for temporal locality. 
In order to derive the temporal CHSH inequality, 'locality in time' is accompanied by an assumption of 'realism' stating that 'the measurement results are determined by ``hidden'' properties the particles carry prior to and independent of observation'. A similar realist constraint is present in the principle of macrorealism. While \citetbjps{adlam2018spooky} warns that these assumptions render irrelevant whether the outcome of a given measurement is correlated in a non-classical way with the outcomes of earlier or later measurements, they constitute at the very least a strong commitment that one might want to avoid for the sake of the full generality of the test. 
 
An alternative definition of temporal locality has been proposed by \citetbjps{adlam2018spooky}, and states in a nutshell that `all influences on a measurement outcome are mediated by the state of the world immediately prior to the measurement'. However, while this definition has the benefit of faithfully capturing the temporal counterpart of Bell locality, it has not yet been used in the context of an experimental setup to derive corresponding temporal inequalities. 
Hopefully, tests involving this formulation of temporal locality could be made free from the above worries affecting the Leggett-Garg and the temporal CHSH inequalities.

The current lack of a satisfying and model-independent (i.e. holding independently of the validity of any particular physical model) test of temporal nonlocality is unfortunate. This prevents us from investigating whether we could elevate this principle from the status of a mere theoretical possibility to that of an observable feature of nature (or at least a feature of physical theories, for the less realist).  

An intuition about the way Adlam's temporal nonlocality could be tested can be gained by examining how temporal correlations are described in quantum formalisms and, if applicable, how one can formulate a corresponding temporal counterpart to standard quantum entanglement. In the same way that entanglement is necessary to Bell nonlocality, one can expect that a suitable concept of temporal entanglement could constitute a necessary ingredient to protocols aiming to test Adlam's temporal nonlocality. 

\textcolor{black}{A relevant work in that regard is the delayed-choice entanglement swapping (DCES) experiment, which was introduced as a thought-experiment by \citetbjps{peres2000delayed}, and was then experimentally realised (see e.g. (\citealtbjps{ma2012experimental}; \citealtbjps{megidish2013entanglement})). These experiments were taken by \citetbjps{glick2019timelike} to involve entanglement between the past and future states of physical systems. Glick claims that the results of the experiments can be explained by allowing a measurement to act backwards in time (which might constitute an instance of temporal nonlocality, as will be discussed in section~\ref{Tloc}), while this ability could be expressed in terms of entanglement relations across time. Objections to this argument have been raised by \citetbjps{egg2013delayed} and \citetbjps{price2021entanglement}, and further developed by \citetbjps{mjelva2024delayed}, who defend a somewhat antirealist take towards these entanglement relations in time. The results of the DCES experiments are explained in terms of a statistical artefact arising due to the presence of a post-selection step in the experimental protocol, at the price of postulating a preferred foliation of spacetime, accepting the relativity of pre- and post-selection, or involving a multiple realisation of outcomes. As \citetbjps{glick2019timelike} concludes, these results remain intriguing and “DCES provides another reason to take timelike entanglement, and hence temporal non-locality, seriously”.} 

\textcolor{black}{If one takes the possibility of temporal entanglement seriously, a central question remains: how should we even formalise entanglement in time?} While quantum entanglement describes correlations among quantum states of distinct systems described at the same time, its temporal counterpart, intuitively, would express correlations among quantum states of systems at different times~\citepbjps{filk2013temporal}. Developing the corresponding mathematical apparatus is non-trivial, as demonstrated in \citepbjps{horsman2017can}. Yet, various proposals for representing quantum states across time and space have been put forward in the literature, including process matrices~\citepbjps{oreshkov2012quantum}, consistent histories~\citepbjps{griffiths1984consistent}, entangled histories~\citepbjps{cotler2016entangled}, quantum-classical games~\citepbjps{gutoski2007toward}, superdensity operators~\citepbjps{cotler2018superdensity}, multi-time states~\citepbjps{aharonov2009multiple}, pseudo-density operators~\citepbjps{fitzsimons2015quantum} and double-density operators~\citepbjps{jia2023spatiotemporal}. Interestingly, \citepbjps{zhang2020quantum} showed that a number of these approaches (namely the process matrix formalism, consistent histories and quantum-classical games) express temporal correlations in a similar, or at least operationally equivalent, way. Within these frameworks, a notion of temporal entanglement has sometimes been proposed, as it is the case for the pseudo-density operators formalism~\citepbjps{ku2018hierarchy} or the notion of entangled histories~(\citealtbjps{nowakowski2017quantum}; \citealtbjps{nowakowski2024entanglement}). In these two specific cases, the role of temporal entanglement in the violation of Leggett-Garg inequalities was discussed. 

It has recently been voiced that indefinite causal orders (defined within the process matrix formalism mentioned above) might constitute another possible notion of temporal entanglement~\citepbjps{crullconf}. This intuition will be put to use in this paper. Our main objective will be to investigate to what extent indefinite causal orders can constitute the core ingredient of an experimental protocol testing Adlam's principle of temporal locality.

The structure of the paper will be as follows. Section~\ref{Tloc} will introduce the definition of temporal locality as proposed by \citetbjps{adlam2018spooky}. Section~\ref{ICO} will present the theory-dependent notion of indefinite causal orders (ICO), as defined in the context of the process matrix formalism. Section~\ref{TnonlocFromICO} will show how the principle of temporal locality can be tested within a setup involving ICOs. More precisely, section~\ref{QS} will present a famous process instantiating an ICO, namely the quantum switch. Section~\ref{setup} will present a setup, proposed in \citepbjps{zych2019bell}, involving two entangled quantum switches, from which 'temporal Bell inequalities' can be derived. Section~\ref{TvsS} will then discuss to what extent temporal nonlocality can be inferred from a violation of these inequalities. Section~\ref{dep} will highlight the model-dependence of that inference, and explore the prospects of a model-independent version of the protocol to test temporal locality. Finally, section~\ref{Dist} will discuss the question of the (physical) distinction between spatial and temporal nonlocality. Section~\ref{ccl} will conclude.  

\section{Temporal Locality}\label{Tloc}

The principle of `local causality' in Bell theorem can be replaced by its straightforward temporal counterpart. Recall that `local causality' means that `causes precede their effects, and causal influences travel continuously through spacetime at subluminal speeds', and can be expressed more rigorously as follows (wording from \citepbjps[p.~1]{adlam2018spooky}):

\begin{quote}
    Suppose that two observers, Alice and Bob, perform measurements on a shared physical system: Alice performs a measurement with setting a and obtains a measurement outcome A, while Bob performs a measurement with measurement setting b and obtains a measurement outcome B. Let $\lambda$ be the joint state of the shared system prior to the two measurements. Then:

    \center $p(A,B|a,b,\lambda)~=~p(A|a,\lambda)~.~p(B|b, \lambda)$
\end{quote}

\begin{figure}
  \begin{minipage}[c]{0.5\textwidth}
    \includegraphics[width=0.8\textwidth]{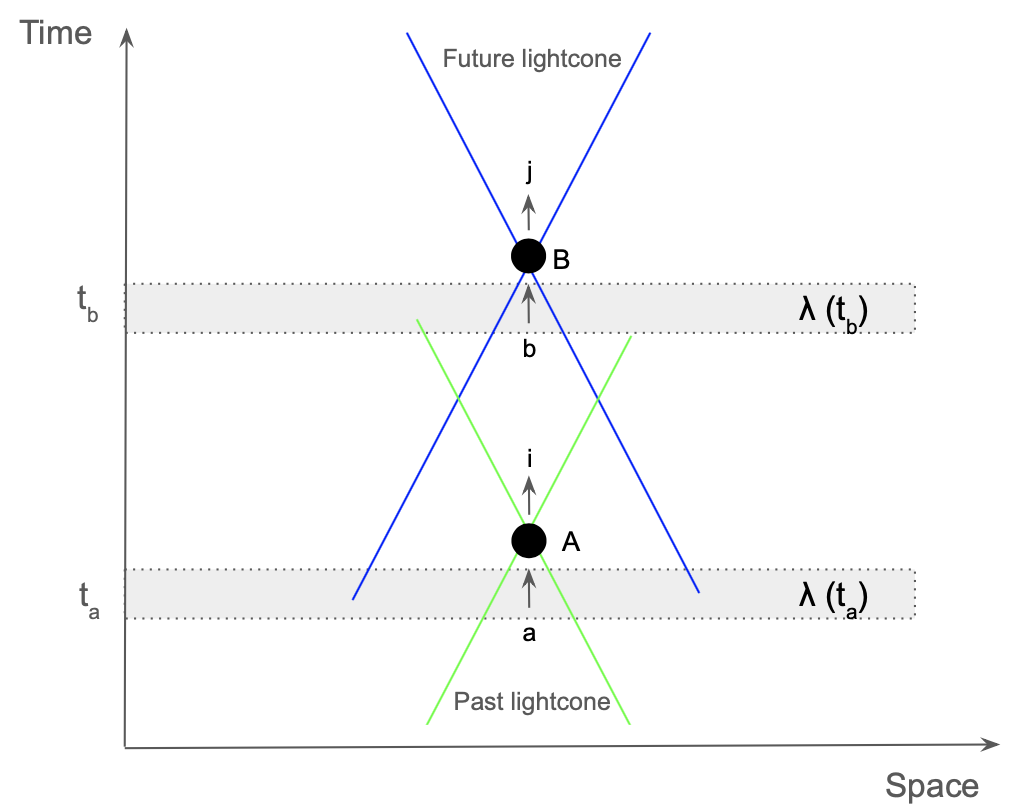}
  \end{minipage}
  \begin{minipage}[c]{0.5\textwidth}
  \caption{Schematic representation of a bipartite scenario with parties A and B having (respectively) their classical inputs denoted a and b, their classical outputs denoted i and j, and the variables $\lambda(t_a)$ and $\lambda(t_b)$ describing the state of the world immediately before their measurement takes place. The present example displays the order $A \preceq B$, yet the definition of temporal locality is valid for any partial order between the parties.}\label{tempdiag}
  \end{minipage}
\end{figure}

Within a given frame of reference specifying a specific time coordinate, \citetbjps[p.~2]{adlam2018spooky} proposed the following temporal version of Bell nonlocality (see Fig.~\ref{tempdiag}): 

\begin{quote}
    Suppose that two observers, Alice and Bob, perform measurements on a shared physical system. At some time $t_a$, Alice performs a measurement with measurement setting $a$ and at some time $t_a + \delta$ she obtains a measurement outcome $i$; likewise, at some time $t_b$, Bob performs a measurement with measurement setting $b$ and at some time $t_b + \delta$ he obtains a measurement outcome $j$. Let $\lambda(t_a)$ be the state of the world at time $t_a$ and let $\lambda(t_b)$ be the state of the world at time $t_b$. Then, temporal locality means that the following equalities hold:
\begin{equation}
    p(i|a,b,\lambda(t_a), \lambda(t_b), j)~=~p(i|a,\lambda(t_a))
    \label{a}
\end{equation}
\begin{equation}
    p(j|a,b,\lambda(t_a), \lambda(t_b), i)~=~p(j|b,\lambda(t_b))
    \label{b}
\end{equation}
\noindent Eq.~\eqref{a} and \eqref{b} can then be combined to yield: 

\center $p(i,j|a,b,\lambda(t_a), \lambda(t_b))~=~p(i|a,\lambda(t_a))~.~p(j|b, \lambda(t_b))$

\end{quote}

Because temporal (non)locality characterises correlations between physical events, this criterion should be frame-independent. Yet, the notion of the ‘state of the world at a given time’ is both frame-dependent and at the centre of the definition of temporal locality. It therefore makes sense that the criterion for temporal locality should (implicitly) hold in any given frame of reference, if a correlation is to be qualified as temporally local.
%\footnote{\textcolor{green}{This is actually similar to the case of spatial (non)locality. A nonlocal correlation between a pair of spacelike separated events can be given a causal story in a given reference frame in which they are non-simultaneous. However, this causal story will be incompatible with the story told in in frame of references for which the order between the events is reversed. Finally, it is not even available in the frame of reference in which these events are simultaneous. As such, we also have the idea that locality imposes that a classical causal story can be told in \textit{any} reference frames.}}

The above principle can be summarised in the following way: `[...] all influences on a measurement outcome would be mediated by the state of the world immediately prior to the measurement'~\citepbjps[p.~2]{adlam2018spooky}. This definition will be adopted in the remainder of this work as the principle of temporal locality to be tested. 

\color{black}
In \citepbjps{adlam2018spooky}, it is proposed that violating temporal locality as defined above can occur if an explicitly retrocausal dynamics is involved, or if the dynamics is said to be 'atemporal' (i.e. the events are determined 'all-at-once' in a global fashion across spacetime), or non-Markovian. However, an important precision is needed. Only the retrocausal influences that act on the measurement's outcome of an earlier party would necessarily violate Eq.~\eqref{a}. One can indeed imagine, in Fig.~\ref{tempdiag}, that a retrocausal influence of party B on party A is encoded in the variable $\lambda(t_a)$, and no violation of Eq.~\eqref{a} would be observed. Retrocausal theories, as long as they do not violate Eq.~\eqref{a} or Eq.~\eqref{b}, would not, strictly speaking, be considered as temporally nonlocal according to Adlam’s definition. A similar reasoning holds for the other candidate mechanisms possibly yielding temporal nonlocality.

%First of all, one might look at continuous backward influences as local phenomena. One’s stance on this question might depend on whether one conceives primarily nonlocality as a discontinuous influence (as captured by the notion of ‘action-at-a-distance’), or as a phenomenon incompatible with local causality more globally (as presented in the second formulation of Bell’s theorem). Proponents of the latter case (in the spirit of Adlam's definition of temporal locality) would argue that, since retrocausality implies a rejection of the principle of causality (and, according to some, of the principle of relativity itself \citepbjps{gao2022retrocausal}), it would violate relativistic causality, hence, local causality.
%However, it is a known issue that retrocausal influences are not operationally detectable unless they are of the signalling kind, while there exists no good evidence supporting the existence of the latter in nature~\citepbjps[section~7.3]{sep-qm-retrocausality}. This problem remains in the context of Adlam's definition of temporal locality: 

%The stance of this paper is the following: from a \textit{conceptual} point of view, non-signalling retrocausal influences, while undetectable operationally, should be seen as a violation of temporal locality, itself defined in terms of the satisfaction of local causality between timelike separated events. Such an instance of temporal nonlocality could still be \textit{inferred} (for lack of an unambiguous operational detection) modulo supporting assumptions. Such considerations will be further discussed in section~\ref{dep}.
    
\color{black}
\section{Indefinite Causal Orders}\label{ICO}

\begin{figure}[H]
    \centering
    \includegraphics[width=0.9\textwidth]{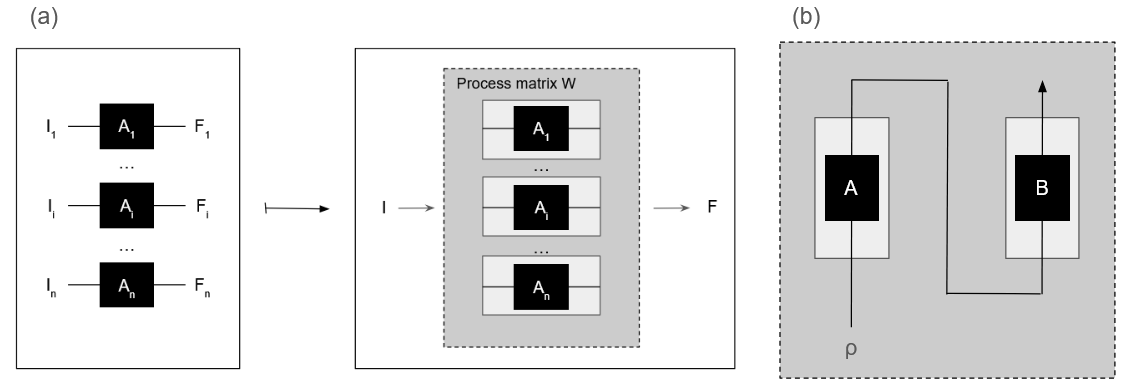}
    \caption{(a) A process matrix $W$ representing a (quantum) process as a map applying $n$ local (quantum) operations (denoted $A_j$ with j going from 1 to $n$) over a global operation. The labels $I$ and $F$ represent the input and output of the process, respectively, while $I_j$ and $F_j$ represent the input and output of the local operation $A_j$, respectively. (b) Process describing one and the same system (initially described by the quantum state $\rho$) undergoing two successive operations ($A$ and $B$), while the final output system is thrown away.}
    \label{processes}
    \end{figure}

In standard quantum mechanics, one can describe correlations (e.g. quantum entanglement) among the quantum states of physical systems. It is possible to zoom out and study quantum correlations at a meta-level, namely among the quantum operations applied to quantum systems. This is achievable within the 'process matrix formalism', of which the central objects are 'processes' represented by 'process matrices' (denoted by $W$). A process is made of a set of parties operating in their respective laboratories, and choosing to perform (on a physical system) some local operation allowed by quantum mechanics. No assumption is made regarding the spatiotemporal locations of the parties. However, it will be described how the various local operations are combined into a global map (see Fig.~\ref{processes} (a)). In that sense, a process will treat the parties as black boxes implementing some local operation, and will describe the quantum dynamics between the parties. Crucially, a valid process will satisfy certain mathematical properties ensuring that the dynamics connecting the parties (even exotic ones, as we will see below) do not give rise to logical contradictions, no matter the local operation that is implemented in each party. 

As an example, Fig.~\ref{processes} (b) shows a bipartite process describing how the quantum input and output systems of two local operations (performed by two isolated parties $A$ and $B$) can be connected to each other.  
  
In the previous example, the order between the local operations is definite. This means that a relation of partial order exists between the involved parties (namely either $A \preceq B$, or $B \preceq A$, or $B \parallel A$), such that the probabilities of the measurements' outcomes of one party can be statistically dependent on the settings of measurements performed by the other party only if this other party is preceding in the partial order. Alternatively, a process can correspond to a probabilistic mixture of partial orders. In that case, it is still the case that the order between the parties is definite.
%the joined probability distributions over the measurements' outcomes of the different parties generated by the whole process 

Let's restate the above definition of definite order. Let $W^{A,B}$ be a bipartite process describing the way two local quantum operations performed on a quantum system by their respective party ($A$ and $B$) are combined to form a global operation. $W^{A,B}$ has a definite order if it can be decomposed as a probabilistic mixture of one-way (or no-) signaling processes (\citealtbjps{oreshkov2012quantum}; \citealtbjps{oreshkov2016causal}):

\begin{equation}
W^{A,B} = q W^{A\prec B} + (1-q) W^{B\prec A}   
\label{causalsep}
\end{equation}
where $q$ is a number between 0 and 1 and $W^{X\prec Y}$ represents a process for which signalling (i.e. the influence of the classical output of a party by the choice of the classical input of another party) is only possible from $X$ to $Y$. 
A generalisation of Eq.~\eqref{causalsep} for multipartite processes has been developed in \citepbjps{oreshkov2016causal} and \citepbjps{wechs2018definition}.

A process with a definite order is also said to be 'causally separable'. Now, because the process matrix formalism only imposes that no logical contradiction arises when the process generates a joined probability distribution over the measurements' outcomes of the different parties, the formalism does not presuppose that the partial order between the parties is definite. 'Indefinite causal order' (ICO) between parties occurs in processes that cannot be decomposed as in Eq.\eqref{causalsep}. These processes are called causally nonseparable. Crucially, there exist causally nonseparable processes that are valid, i.e. describe dynamics between parties that are free from logical contradictions. A concrete example of indefinite causal order, displayed in a process called the quantum switch, is presented in the next section. 

The intuition that causal nonseparability (or, equivalently, indefinite causal orders) is intimately connected to a form of temporal entanglement has been formulated in the literature~\citepbjps{crullconf}.
In this paper, we will put this intuition to use and show that, at the very least, a process involving an ICO can yield a violation of Adlam's principle of temporal locality.

\section{Temporal Nonlocality from Indefinite Causal Orders}\label{TnonlocFromICO}

This section will proceed as follows. First, section~\ref{QS} will present the quantum switch~\citepbjps{chiribella2013quantum}, a process displaying an indefinite causal order that can be physically implemented~(\citealtbjps{procopio2015experimental}; \citealtbjps{rubino2017experimental};  \citealtbjps{goswami2018indefinite};  \citealtbjps{wei2019experimental};  \citealtbjps{guo2020experimental}). From there, section~\ref{setup} will present a more general experimental setup proposed in \citepbjps{zych2019bell}, involving two quantum switches, from which 'temporal Bell inequalities' can be derived. In section~\ref{TvsS}, we will show that the violation of these inequalities implies temporal nonlocality, \textcolor{black}{under certain (model-dependent) assumptions. Section~\ref{dep} will explore the prospects of a model-independent version of the protocol to test temporal locality.}

    \subsection{A case of Indefinite Causal Order: the quantum switch}\label{QS}

Let's consider a particular example of causally nonseparable process, called the quantum switch (QS). This process involves three parties. First, two parties, $A$ and $B$, perform a quantum operation (denoted $U_A$ and $U_B$, respectively) on a target system (of which the quantum state is denoted $\ket{\psi_{t}}$).\footnote{The following assumes unitary operations for simplicity, but one can generalise the description to non-unitary transformations.} The state of a qubit system, called the control (of which the quantum state is denoted $\ket{\psi_{c}}$) determines the order of the operations performed by parties $A$ and $B$: if $\ket{\psi_{c}} = \ket{0_c}$, then the output system of party $A$ is sent to the input of party $B$ (this is denoted $A \preceq B$). If $\ket{\psi_{c}} = \ket{1_c}$, then the output system of party $B$ is sent to the input of party $A$ (this is denoted $B \preceq A$). In the quantum switch, the control system is in a superposition of states, i.e. $\ket{\psi_{c}} = \frac{1}{\sqrt{2}}(\ket{0_c} + \ket{1_c})$. As a result, the order of the operations of parties $A$ and $B$ is entangled with the state of the control, hence indefinite. The final state of the composite system [target + control] (denoted $\ket{\psi^{QS}}$) is expressed as follows: 

\begin{equation}
        \ket{\psi^{QS}} = \frac{1}{\sqrt{2}} (\ket{0_c} U_B U_A \ket{\psi_{t_0}} + \ket{1_c} U_A U_B \ket{\psi_{t_0}})
        \label{QS-final-EQ}
\end{equation}
where $\ket{\psi_{t_0}}$ denotes the state of the target system at the very beginning of the process. A third party, $Z$, receives $\ket{\psi^{QS}}$, and performs some projective measurement on the control system. 

For the sake of completeness (and we will see that it matters in the next section), we should also take into account the (possible) presence of free evolution of the target system in between the operations taking place in parties $A$ and $B$. Let's denote $V_{0_c}$ the operator for the free evolution of the target in case the control system is in the state $\ket{0_{c}}$, and $V_{1_c}$ when the control is in $\ket{1_c}$. Eq.~\eqref{QS-final-EQ} becomes: 

\begin{equation}
        \ket{\psi^{QS}} = \frac{1}{\sqrt{2}} (\ket{0_c} U_B V_{0_c} U_A \ket{\psi_{t_0}} + \ket{1_c} U_A V_{1_c} U_B \ket{\psi_{t_0}})
        \label{QS-final-EQ-free}
\end{equation}

\begin{figure}
  \begin{minipage}[c]{0.5\textwidth}
    \includegraphics[width=0.8\textwidth]{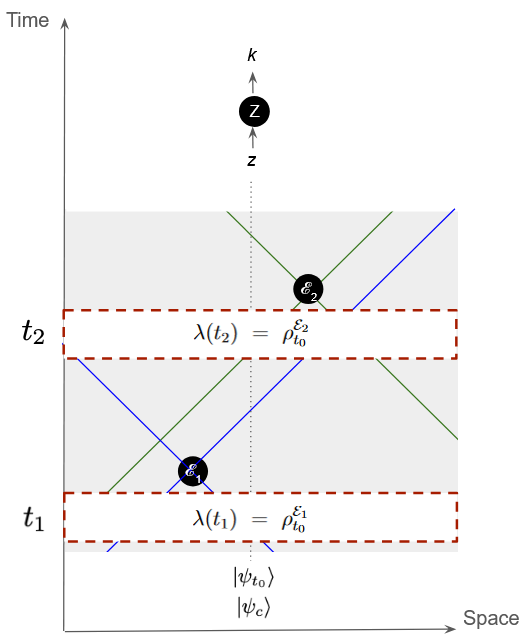}
  \end{minipage}
  \begin{minipage}[c]{0.5\textwidth}
    \caption{Schematic representation of a quantum switch. A preparation procedure prepares the target system in state $\ket{\psi_{t_0}}$ and the control system in state $\ket{\psi_{c}}~=~\frac{1}{2}[\ket{0_c}+\ket{1_c}]$. The target system undergoes an operation in party $A$, according to the map $U_{A}$, and another operation in party $B$, according to the map $U_{B}$. The order between these operations is indefinite. These occurrences of the operations are denoted $\mathcal{E}_1$ and $\mathcal{E}_2$, without specifying to which party they correspond. The states $\rho_{t_0}^{\mathcal{E}_1}$ and $\rho_{t_0}^{\mathcal{E}_2}$ correspond to that of the target system immediately before events $\mathcal{E}_1$ and $\mathcal{E}_2$, respectively. At the end of the process, party $Z$ receives the final state of the composite system [target + control] and makes a projective measurement (with classical setting $z$) on the control system, yielding some classical outcome $k$.} \label{ProtocolQS}
  \end{minipage}
\end{figure}

Fig.~\ref{ProtocolQS} summarises the experimental protocol and the notations introduced above. For the sake of visual representation, we denote $\mathcal{E}_1$ and $\mathcal{E}_2$ the occurring of the transformations undergone by the target system. Since the order between party $A$ and $B$ is indefinite, it is indefinite whether $\mathcal{E}_1$ corresponds to $U_{A}$ or $U_{B}$, and similarly for $\mathcal{E}_2$. The variables $\lambda_(t_1)$ and $\lambda_(t_2)$ encode the state of (the relevant part of) the world immediately before events $\mathcal{E}_1$ and $\mathcal{E}_2$, respectively. By definition of the QS, one sees that $\lambda({t_1})~=~\rho_{t_0}^{\mathcal{E}_1}$ and $\lambda({t_2})~=~\rho_{t_0}^{\mathcal{E}_2}$, i.e. the variables $\lambda({t_1})$ and $\lambda({t_2})$ correspond to the state of the target system immediately before it undergoes event $\mathcal{E}_1$ and $\mathcal{E}_2$, respectively. The grey square on Fig.~\ref{ProtocolQS} represents the spacetime region in which lies the indefinite causal order between parties $A$ and $B$. As explained in \citepbjps{de2022quantum}, it can originate either from a superposition of worldlines of the target system (one worldline would undergo operations $U_{A}$ before $U_{B}$, and the other worldline would do the opposite), or from a superposition of spacetime metrics within that region (one metric would be such that $A \preceq B$, and the other metric would describe the opposite order). 

We will now see, in the next section, how \citetbjps{zych2019bell} devised a more complex setup involving two quantum switches that can be used to derive 'temporal' Bell inequalities.

    \subsection{Experimental setup for testing temporal nonlocality}\label{setup}

\begin{figure}
  \begin{minipage}[c]{0.5\textwidth}
    \includegraphics[width=0.8\textwidth]{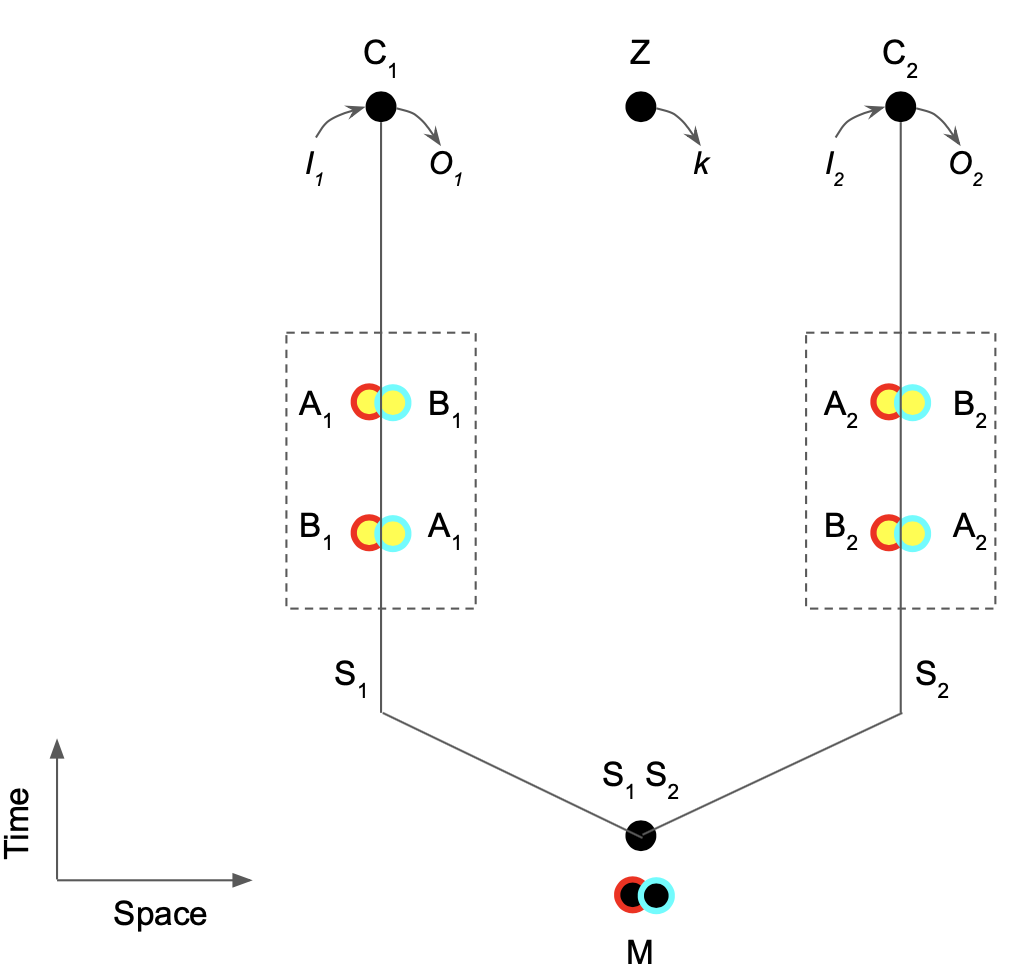}
  \end{minipage}
  \begin{minipage}[c]{0.5\textwidth}
    \caption{Schematics of the setup for the violation of a temporal Bell inequality, as proposed in \citepbjps{zych2019bell}. A massive body $M$ is in a superposition of two mass configurations $\ket{M^{A \preceq B}}$ and $\ket{M^{B \preceq A}}$, each inducing the partial order $A \preceq B$ and $B \preceq A$, respectively. Two target systems $S_1$ and $S_2$ are prepared in a product state $\ket{S_1}\ket{S_2}$. $S_1$ is sent to a first quantum switch, while $S_2$ is sent to a second one, spacelike-separated from the first. Each party $A_i$ and $B_i$ (i = 1,2) in each quantum switch performs one operation. The order between these operations is indefinite. Two spacelike-separated parties ($C_1$ and $C_2$) measure $S_1$ and $S_2$, respectively. The input (output) of party $C_i$ is denoted $I_i$ ($O_i$). The control system $M$ is measured by party $Z$ (in the future lightcones of parties $A_i$ and $B_i$ (i = 1,2), and spacelike separated from $C_1$ and $C_2$), producing an output bit $k$.} \label{zych5}
  \end{minipage}
\end{figure}

The setup of \citetbjps{zych2019bell} (see Fig.~\ref{zych5}) involves two quantum switches relying on gravitational effects to induce the ICO. The control system $M$ is a massive body in a superposition of two mass configurations $\ket{M^{A \preceq B}}$ and $\ket{M^{B \preceq A}}$. Each of these states induces, via an appropriate curvature of spacetime, a specific order between the operations of parties $A$ and $B$, namely $A \preceq B$ and $B \preceq A$, respectively. Two such quantum switches are spacelike-separated from each other. Two target systems, $S_1$ and $S_2$, are prepared in a product state $\ket{S_1}\ket{S_2}$. $S_1$ is sent to the first quantum switch, while $S_2$ is sent to the second. After these systems underwent operations in parties $A_i$ and $B_i$ (i = 1,2), two spacelike-separated parties ($C_1$ and $C_2$) perform a measurement on systems $S_1$ and $S_2$, respectively. The classical input (output) of party $C_i$ is denoted $I_i$ ($O_i$). The control system $M$ is measured by another party $Z$ (in the future lightcones of parties $A_i$ and $B_i$ (i = 1,2), and spacelike separated from $C_1$ and $C_2$), producing an output bit $k$.

\citetbjps{zych2019bell} derived a Bell inequality, characterising the correlation between the spacelike-separated events at $C_1$ and $C_2$. The following premises have been assumed for the derivation: 

\begin{itemize}
\setlength\itemsep{0.01em}
    \item [A1] The initial state of the two target systems and the control system is separable. By definition, a state $w$ is separable (written as $w = w_1 \otimes w_2$), if the probabilities of local measurements factorise as \textcolor{black}{$P(O_1, O_2 |I_1 , I_2 , w) = P(O_1 |I_1 , w_1 ).P(O_2 |I_2 , w_2)$}, with $O_i$ and $I_i$ being the classical outputs and inputs of the measurements. 
    
    \item [A2] The operations performed on the systems are local, i.e. they are realised at the time and location defined by a local clock.

    \item [A3] The parties $A_1$ and $B_1$ are space-like-separated from the parties $A_2$ and $B_2$. The parties $C_1$, $C_2$ and $Z$ are pair-wise spacelike-separated.

    \item [A4] The choice of the settings of the measurements in $C_1$, $C_2$ and $Z$ is statistically independent from the rest of the experiment. 

    \item [A5] Any non-trivial free evolution of the target system $S_i$ in between parties $A_i$ and $B_i$ remains uncorrelated to the control system's state. 

    \item [A6] The parties $A_i$ and $B_i$ are 'classically ordered', i.e. `there exists a space-like surface and a classical variable $\gamma$ defined on it that determines the causal relation between $A_i$ and $B_i$: either $A_i \preceq B_i$ ($A_i$ is in the past causal cone of $B_i$), $B_i\preceq A_i$ ($A_i$ is in the past causal cone of $B_i$) or $A_i ||B_i $ ($A_i$ and $B_i$ are space-like separated)'~\citepbjps[p.~4]{zych2019bell}
    
\end{itemize}

As a note, the assumption [A5] regarding the free evolution of the target in between parties $A_i$ and $B_i$ was not explicitly present in the original work of \citetbjps{zych2019bell}, and has been made explicit in \citepbjps{dkebski2022indefinite}.\footnote{The discussion in \citepbjps{dkebski2022indefinite} is based on a variant experimental setup in which the quantum switches are not implemented via a superposition of mass configurations, but via relativistic time dilation effects. However, their argument showing that assumption [A5] is assumed in the derivation of the temporal Bell inequality is valid for the gravitational version as well.} As it will be discussed in section~\ref{dep}, the satisfaction of this assumption is arguably not verifiable in a model-independent way.

The authors then show that the joint final state of the target systems $S_1$ and $S_2$ (denoted $\ket{\psi_{final}}$) is entangled, which means that there exist local measurements that can be performed on them by parties $C_1$ and $C_2$ such that the corresponding Bell inequality is violated. This joint final state, displaying two entangled QS, has a form similar to Eq.~\eqref{QS-final-EQ-free}: 

\begin{equation}
    \begin{split}
        \ket{\psi_{final}} = \frac{1}{\sqrt{2}} &(\ket{M^{A \preceq B}} U_{B_1} V_{1}^{AB} U_{A_1} \ket{\psi_{S_1}} U_{B_2} V_{2}^{AB} U_{A_2} \ket{\psi_{S_2}} \\
        &+ \ket{M^{B \preceq A}} U_{A_1} V_{1}^{BA} U_{B_1} \ket{\psi_{S_1}} U_{A_2}V_{2}^{BA} U_{B_2} \ket{\psi_{S_2}})
          \end{split}
           \label{QS-final-EQ-zych}
\end{equation}

where $V_{i}^{XY}$ is the operator describing the free evolution of the target system $S_i$ in the branch with causal order $X \preceq Y$. It follows from this result that at least one of the assumptions [A1-A6] needs to be rejected.

    \subsection{Temporal versus spatial nonlocality}\label{TvsS}

Assumptions [A1-A4] are arguably reasonable, although some could defend, e.g., a rejection of the free choice of settings, as it has been done in the context of Bell theorem~(\citealtbjps{hossenfelder2020rethinking}; \citealtbjps{andreoletti2022superdeterminism}; \citealtbjps{baas2023does}). However, if one deems [A1-A4] acceptable, in the face of the above-mentioned violation of the temporal Bell inequality, which premise, of [A5] or [A6], should we reject? 

Let's assume that the assumption [A5], namely the guarantee that any non-trivial free evolution of the target remains uncorrelated to the control system, is violated, while assumption [A6] holds. In that case, there would exist a variable $\gamma$ on a spacelike surface such that parties $A_i$ and $B_i$ enter a definite partial order relation. Yet, Eq.~\eqref{QS-final-EQ-zych} would still display an entangled state due to the difference in the overall evolution of the target system (including its free evolution, that is) depending on the state of the control. One could imagine that instead of the proper implementation of the setup in Fig.~\ref{zych5}, the mass configuration would only induce a different spacetime metric in between the events occurring in $A_i$ and $B_i$, while these parties would lie in a definite causal order. 

In that case, and as mentioned in section~\ref{intro}, it is always possible to reject the principle of temporal locality to explain the presence of the Bell nonlocal correlation between the spacelike-separated events in $C_1$ and $C_2$. Yet, it would not be necessary to appeal to its violation to account for the final entangled state in Eq.~\eqref{QS-final-EQ-zych}. The situation, at least as it is modelled with the tools of section~\ref{ICO}, is compatible with the idea that whatever happens in $A_i$ and $B_i$ depends only on the state of the world immediately prior to their operations. To see this, let's go back to Fig.~\ref{ProtocolQS}, in which we would consider the order between parties $A_i$ and $B_i$ as fixed (corresponding to, say, $A_i \preceq B_i$, with $\mathcal{E}_1$ ($\mathcal{E}_2$) corresponding to party $A_i$ ($B_i$)).  We would have that the state $\rho_{t_0}^{\mathcal{E}_2}$ (i.e. $\rho_{t_0}^{B_i}$) would be mixed (following the indeterminacy of the target's free evolution between party $A_i$ and $B_i$), yet it would contain (and therefore screen off) any contribution from party $A_i$, such as $\rho_{t_0}$ and the setting and outcome of the measurement performed on the target by $A_i$. Reciprocally, the model does not explicitly show a retro-influence of party $B_i$ on party $A_i$, as the state $\rho_{t_0}^{\mathcal{E}_1}$ (i.e. $\rho_{t_0}^{A_i}$) is definite (if $\rho_{t_0}$ is so) and does not contain explicit information about party $B_i$. 

\color{black}
Let’s suppose now that the assumption [A6], namely that of classical order, is violated, while assumption [A5] holds (for any possible theoretical model describing the free evolution). 

\begin{enumerate}
\setlength\itemsep{0.01em}
    \item In the demonstration of \citetbjps{zych2019bell}’s theorem, the assumption of classical order [A6] is taken to imply that, if the order between parties $A_i$ and $B_i$ is $A_i < B_i $, then this means that the operation of party $A_i$ occurs in the past lightcone of party $B_i$. As a result, the operation of $A_i$ can, possibly, causally influence party $B_i$, but not the contrary. A similar logic applies for the order $B_i < A_i $. If $A_i || B_i $, then no party can causally influence the other one. 
    %\item The possibility of signalling implies the possibility of exerting some causal influence. 
    \item If [A6] is violated, there exists no variable $\gamma$ on a spacelike surface such that a partial order exists between parties $A_i$ and $B_i$ (i.e. either $A_i  \preceq B_i $, $B_i \preceq A_i$ or $A_i||B_i$).\footnote{As shown in \citepbjps{de2022quantum}, an ICO is a frame-independent feature, and cannot be dispelled upon selecting a particular frame of reference. The presence of an ICO is therefore consistent with the idea that, in spatiotemporal terms, there exists no variable $\gamma$ on a spacelike surface for which a definite causal order exists.} 
    \item From the two above claims, the following ensues: since the experiment is incompatible with $A_i ||B_i $, it is false that neither party $A_i$ nor party $B_i$ exerts some causal influence on the other party. In other words, the fact that no model can describe $A_i$ and $B_i$ being spacelike separated means that there must be some causal influence between the two parties. 
    %\item Yet, bi-directional signalling between parties $A_i$ and $B_i$ does not take place in the QS, as demonstrated in the literature (see, e.g., (see, e.g., \citepbjps{araujo2015witnessing, oreshkov2016causal, van2023device}).
    %\item Therefore, signalling must be present alongside a unique direction, namely either from party $A_i$ to party $B_i$, or reciprocally. 
    \item If only party $A_i$ causally influences party $B_i$, then a model according to which the order $A_i < B_i$ holds should be adequately describing the setup (and reciprocally for causal influences from party $B_i$ to party $A_i$).
    \item However, violation of [A6] means that neither $A_i < B_i $ nor $B_i < A_i $ is a compatible description, no matter what the model and its variable $\gamma$ are. 
    \item It ensues that causal influences must be happening in both directions of time, which amounts to accepting retrocausal influences. 
\end{enumerate}
%no need for non-commutativity of operations here ? no, in the statistics of the QS, we change the local operations (i.e. change the settings), so different combinations arise. 

As explained in section~\ref{Tloc}, these retrocausal influences will imply temporal nonlocality if they retroactively impact the measurement's outcome of earlier parties. This last step in the argument is discussed below. 

The reading of the QS as involving retrocausal influences is a valid understanding of the process. The work of \citetbjps{vilasini2022embedding} demonstrated that a process matrix (which maps local operations into a global one) can be alternatively described as a composition of operations (possibly) involving feedback loops (i.e. an output system of an operation is looped back and connected to an input system of that same operation). Importantly, \citetbjps[corollary~6.3]{vilasini2022embedding} proved, among more general claims, that a process with an ICO is such that the structure connecting its (information-theoretical, i.e. not embedded in spacetime) input and output systems (called information-theoretic causal structure) is incompatible with an acyclic directed graph. Yet, there exists a compatible cyclic causal structure.\footnote{This result generalises the work of \citetbjps{barrett2021cyclic}, which demonstrated that unitary processes with ICOs are equivalent to processes in which the order among parties is definite, yet cyclic.} This sheds light on the range of possible physical realities underlying ICOs. As described within the PMF, the incompatibility of a process with any acyclic (information-theoretic) causal order is taken to indicate that the causal structure is indefinite. However, as equivalently modelled in the framework of \citetbjps{vilasini2022embedding}, this incompatibility can be taken to indicate that the causal structure is definite, yet cyclic. 

If one appeals to the latter model, provided that this cyclic 'information-theoretic' causal structure translates into a cyclic 'relativistic' causal structure upon embedding the process in spacetime (see section~\ref{dep} for a discussion), this cyclic relativistic causal structure would allow for retrocausal influences between well-defined spatiotemporal events (namely the spatiotemporal embedding of the quantum input and output systems of parties $A_i$ and $B_i$ in the setup of Fig.~\ref{zych5}). Since no alternative acyclic relativistic causal structure would be compatible with the experiment's behaviour, we can infer that retrocausal influences would indeed be taking place.

\textcolor{black}{Importantly, \citetbjps{vilasini2022embedding}'s framework considers feedback loops such that, in the case of the QS, the quantum outputs of each party causally influence the inputs of the other party. This model would yield no violation of the principle of temporal locality for reasons explained in section~\ref{Tloc}. However, a theoretical model for which these retrocausal influences impact the measurement's outputs of earlier parties (denoted 'output-impacting' retrocausality from now on), instead of their input, is conceivable and could be posited with the same experimental predictions.\footnote{\textcolor{black}{Major retrocausal models found in the literature are explicitly input-impacting (e.g., the models in (\citealtbjps{schulman1997time}; \citealtbjps{schulman2012experimental})). In the two-state vector formalism (TSVF) (see, e.g., \citepbjps{aharonov2008two}) the probabilities of obtaining a given outcome for a measurement is explicitly dependent on future events (see Eq. (13.9) and (13.10) in \citepbjps{aharonov2008two}), but this dependency seems to be the sole consequence of having the inputs of the measurement dependent on future events. An output-impacting retrocausal model could have that the classical settings of a measurement be impacted by future events, so that the output depends on future events via other means than the mere retrocausal influence on the input. To the best of my knowledge, such a scenario has not been discussed in the literature, and would correspond to a kind of “time-symmetric superdeterminism”. I leave it as an open question whether such a proposal would display any virtue (e.g., while time-symmetry can be considered as an attractive model feature for conceptual or philosophical reasons, some could see the combination of superdeterminism with retrocausality as a weakening move~\citepbjps{hossenfelder2020superdeterminism}.}} 
One sees, however, that this further constraint makes the presence of temporal nonlocality from retrocausality highly model-dependent. Yet, this situation might be understood as an artefact of a possibly too restrictive definition of temporal locality. Indeed, a slight modification of Adlam's original definition would encompass both 'output-impacting' and 'input-impacting' retrocausal models as possible instances of temporal nonlocality, without significantly altering the physical meaning of the latter.}

\textcolor{black}{One might want to define the variables $\lambda (t_A)$ and $\lambda (t_B)$ in Eq.~\eqref{a} and Eq.~\eqref{b}, not as `descriptions of the world immediately prior to the measurements', but simply as `descriptions of the world prior to the measurements'. 
The fact that these descriptions can correspond to arbitrarily remote regions of the past ensures that temporal locality can be violated in the case of 'input-impacting' retrocausal models. The idea is that, while some variables $\lambda (t_A)$ and $\lambda (t_B)$ characterising some input for the measurements may be impacted by future events, and would not lead to a violation of temporal locality if used for the calculations of the outcomes' probabilities, there could exist some other set of variables $\lambda (t'_A)$ and $\lambda (t'_B)$ (with $t'_X < t_X,~ X=A, B$) lying in the past of these retrocausally modified inputs, for which no influence from the future is taking place. Appealing to this set of variables for the calculation of the measurements' predictions would then amount to seeing the backward modification of the outputs of measurements (following the backward modifications of the variables $\lambda (t_A)$ and $\lambda (t_B)$ characterising their input), which would violate temporal locality. An example of such a situation in the case of the QS would be to take the variables $\lambda (t'_A)$ and $\lambda (t'_B)$ to be the initial state of the target system as it enters the process ($\ket{\psi_{t_0}}$), while $\lambda (t_A)$ and $\lambda (t_B)$ would correspond to the target system's mixed states at the entrance of parties A and B, respectively. While $\lambda (t_A)$ and $\lambda (t_B)$ are influenced by the events taking place in the other party, it is not the case of $\ket{\psi_{t_0}}$, which is not affected by what takes place in the process.}
%It is rather straightforward to show that, for the QS expressed in the framework of \citetbjps{vilasini2022embedding}, one can violate $p(i|a,b,\ket{\psi_{t_0}}, j)~=~p(i|a,\ket{\psi_{t_0}})$.
%
%\footnote{Note that \textit{complete} here means that the descriptions includes all that is relevant for the calculation of probabilities \textit{according to the theory upon consideration}. No assumption is made regarding whether the theory itself is complete~\citepbjps{norsen2011john}.} 

\textcolor{black}{This slight change from temporal locality as the idea that “[...] all influences on a
measurement outcome would be mediated by the state of the world \textit{immediately prior to} the measurement”~\citepbjps[emphasis added]{adlam2018spooky}, to temporal locality as the idea that "all influences on a
measurement outcome would be mediated by the state of the world \textit{prior to} the measurement" would not imply a significant departure from the original definition, and would remain in close agreement with the initial conception of Bell's local causality (here applied to timelike-separated events) as the requirement that “causes precede their effects, and
causal influences travel continuously through spacetime at subluminal speeds”. I leave open, at this stage, the question of whether such a variant should be preferred. For the time being, we adopt the model in which retrocausal influences impact the measurement's outcomes of earlier parties, from which a violation of temporal locality can be inferred.} 

\color{black}
We already see how the inference of temporal nonlocality from a violation of a temporal Bell inequality relies on model-dependent assumptions, namely a particular modelisation of ICOs in terms of a definite, yet cyclic relativistic causal structure. More generally, the above argument does not even refer to actual measurements that would be made in parties $A_i$ and $B_i$. Indeed, temporal nonlocality is inferred from the correlation between measurements performed in parties $C_1$ and $C_2$. Strictly speaking, the measurements in parties $A_i$ and $B_i$ are merely assumed to take place. We will see that whether this commitment holds will depend both on the spatiotemporal embedding of the QS's implementation, and on the physical interpretation of ICOs themselves. Another weakness of the above argument is that we have assumed that a violation of a temporal Bell inequality would point towards a violation of the principle of classical order, rather than a violation of assumption [A5]. The next section will discuss in more detail whether the present test of temporal locality can be made model-independent in spite of (i) relying on a rejection of the principle of classical order instead of a rejection of assumption [A5], and (ii) assuming an interpretation of the QS as involving retrocausal influences between operations that actually take place. 

\color{black}

    \subsection{A model-independent test?}\label{dep}

In the face of a violation of a temporal Bell inequality when implementing the setup of \citetbjps{zych2019bell} (and setting aside the assumptions [A1-A4]), one would need convincing reasons to claim that [A5] holds. Yet, settling the question of whether [A5] is satisfied would require describing the experimental protocol with a theoretical model. This is therefore a model-dependent statement. This point was already raised in \citepbjps{dkebski2022indefinite}. 

\textcolor{black}{We note, however, that the above argument inferring temporal nonlocality from a non-classical order remains valid even if assumption [A5] were to be violated simultaneously to [A6]: the presence of a non-trivial free evolution between $A_i$ and $B_i$ that would be coherently controlled would not cancel out the retrocausal influences taking place between the parties. As such, a model-independent guarantee that at least [A6] is violated via (output-impacting) retrocausal influences would be enough for the whole test to be model-independent. This would require an implementation of the QS so that we can guarantee, model-independently, that it instantiates an (output-impacting) cyclic relativistic causal structure among well-localised spacetime events.}

There is an ongoing debate regarding the way we should understand the spatiotemporal embedding of the quantum switch's realisations. We recall that the 'causal order' between two parties of a process is to be understood in information-theoretic terms (whether that causal structure is modelled in the process matrix formalism or in \citetbjps{vilasini2022embedding}'s framework). The connection between the information-theoretic structure and its corresponding spacetime embedding is not trivial~(\citealtbjps{oreshkov2019time}; \citealtbjps{vilasini2022embedding}): how are the physical systems and their interactions embedded in spacetime? Does the information-theoretic ICO translate into an indefinite causal order in terms of spatiotemporal relations between spacetime events? \textcolor{black}{Let's review three different scenarios for the spatiotemporal embedding of the QS, and their implications on our argument.}

\color{black}

\begin{enumerate}
\setlength\itemsep{0.01em}
    \item \textbf{Simulation of ICOs in a fixed spacetime:} Let's assume a fixed spacetime background. When experimentally realising an ICO, the inputs and outputs systems of the different parties become embedded in that fixed spacetime. It might very well be the case that this embedding is such that their location is indefinite. 
    
    In the case of the QS, it would be indeterminate when and/or where the target system undergoes the operations in parties A and B. In particular, the information-theoretic event consisting of the operation of party A on the target system would correspond to a pair of spacetime events ($\mathcal{E}^{A \preceq B}_{A}$ and $\mathcal{E}^{B \preceq A}_{A}$), each of them associated to the control system's state $\ket{0^c}$ or $\ket{1^c}$, respectively. Which of these two events obtains is indefinite due to the fact that the control system is in the state $\frac{1}{\sqrt{2}}(\ket{0^c}+\ket{1^c})$. A similar situation would occur for party B. The target would be in a superposition of worldlines, one passing through the pair of events ($\mathcal{E}^{A \preceq B}_{A}$, $\mathcal{E}^{A \preceq B}_{B}$), and the other passing through the pair of events ($\mathcal{E}^{B \preceq A}_{A}$, $\mathcal{E}^{B \preceq A}_{B}$).\footnote{Two examples of such implementations involving spatiotemporal delocalisations of the events in parties A and B can be found in (\citealtbjps{paunkovic2020causal}; \citealtbjps{vilasini2022embedding}).} As discussed in \citepbjps{paunkovic2020causal} and \citepbjps[section~7.4]{vilasini2022embedding}, we would not have a 'proper' ICO in that spacetime embedding of the process, but rather a superposition of relativistic causal orders between different pairs of spatiotemporal events. In that context, it can be shown that a fixed, acyclic, relativistic causal order can be assigned to the set of spatiotemporal events, and it is merely indeterminate which worldline along this fixed causal structure obtains for the target system.

    Importantly, because it would be indefinite which spatiotemporal events occur, one can hardly claim that actual measurements take place in parties A and B. As a result, it becomes meaningless to ask whether the principle of temporal locality is obeyed within the process, since we do not have physically meaningful measurements that can be localised in spacetime and that could possibly yield temporally (non)local correlations. %just like it is meaningless to speak of spatial nonlocality between these events as well ? add? 
    In more detail, we need well-defined events to make sense of the variables $\lambda(t_a)$ and $\lambda(t_b)$ appearing in the definition of temporal locality as presented in section~\ref{Tloc}. The absence, at least in some frames of references, of well-defined variables $\lambda(t_a)$ and $\lambda(t_b)$ renders such a definition inapplicable. Strictly speaking, we would therefore be facing a case of inapplicability of the concept of temporal locality, rather than a proper violation thereof.\footnote{\textcolor{black}{It is worth comparing the tension between temporal locality and relativity on the one hand, and between temporal locality and metaphysical indeterminacy of causal orders on the other hand. While relativity makes the notion of “state of the world at a certain time” frame-dependent, this does not render the definition of temporal locality inapplicable. Instead, it seems to require including the stronger constraint that the corresponding principle should be obeyed in any frame of reference for temporal locality to hold (as discussed in section~\ref{Tloc}). On the other hand, the metaphysical indeterminacy of the causal order renders the notion of temporal locality plainly inapplicable, since it makes it impossible to even define, in at least some frames of reference, a global temporal coordinate to specify the “state of the world at a certain time” before the local operations.}}

    Regarding the experiment of Fig.~\ref{zych5}, the target system in each QS would be in a superposition of worldlines, one in which it passes through the pair of spatiotemporal events ($\mathcal{E}^{A_i \preceq B_i}_{A_i}$, $\mathcal{E}^{A_i \preceq B_i}_{B_i}$), and one in which it passes through a different pair of spatiotemporal events ($\mathcal{E}^{B_i \preceq A_i}_{A_i}$, $\mathcal{E}^{B_i \preceq A_i}_{B_i}$). The violation of a temporal Bell inequality between parties $C_1$ and $C_2$ would follow from a violation of assumption [A5]: the final state in Eq.~\eqref{QS-final-EQ-zych} gets entangled because of the indeterminate overall spatiotemporal evolution of the target depending on the state of the control. 

    \item \textbf{Proper ICO with an 'indefinite spacetime':} Alternatively, the embedding of an ICO can be such that the information-theoretic events are delocalised in spacetime, not because the systems follow indefinite worldlines against a fixed spacetime background, but because spacetime itself is indefinite. 
    In that case, one would have a true superposition of causal orders between the same pair of events, and the information-theoretic ICO would translate into an indefinite relativistic causal structure between two spatiotemporal events.
    
    One can take this description literally from an ontological point of view, as proposed in \citepbjps{letertre2022causal}: metaphysically speaking, there would be no definite relativistic order between the spatiotemporal events. The objective world, and not our knowledge thereof, would be such that this order is indefinite. Distinct theories of metaphysical indeterminacy could be used to articulate more precisely this idea. 
    %going from MI of ICO to MI of relativistic order: check steps. 

    In the context of \citetbjps{zych2019bell}'s experiment, this scenario would amount to take literally the violation of assumption [A6]: there is no variable $\gamma$ on a spacelike surface determining the partial order of the events, because this order is metaphysically indeterminate.\footnote{\textcolor{black}{If the metaphysical indeterminacy of the order between events is postulated to be the underlying physical interpretation of a violation of a temporal Bell inequality in \citetbjps{zych2019bell}'s setup, more work would be required to fledge an appropriate explanation of the phenomenon. A minimal requirement would be to provide an account of laws of nature that do not require the existence of a definite order between events to be articulated. This could be done by appealing to 'global' laws that would apply macroscopic constraints to the entire world taken as a whole, while these constraints would leave some features (e.g. the causal order between certain pairs of events) undetermined at the microscopic level. This proposal is inspired by what has been suggested in \citepbjps[p.~6]{adlam2019fundamental}. Accounts of laws as `global constraints' have been suggested in the recent literature~(\citealtbjps{adlam2022laws}; \citealtbjps{Chen2022}; \citealtbjps{meacham2023nomic}).}} %The notion of classical order between these events might become inapplicable altogether. 

    \textcolor{black}{Embracing the metaphysical indeterminacy of the relativistic order between two spatiotemporal events can affect different aspects of reality, depending on one's specific ontological picture of reality, in particular the ontological account of spacetime and its relation to matter. Without entering such details, one could argue that a metaphysically indeterminate relativistic order between two spatiotemporal events amounts to saying that the events themselves are not well-defined. Similarly to the case of simulations of ICOs in a fixed spacetime, this threatens to render the concept of temporal (non)locality inapplicable.}
    
    \item \textbf{Proper ICO with a cyclic structure}: Let's assume that the information-theoretic inputs and outputs in the process involving an ICO are well localised in spacetime. In that case, the cyclic (information-theoretic) causal structure between the inputs and outputs of the local operations within the process would get translated into a definite, yet cyclic, relativistic causal structure between spatiotemporal events. In the case of the QS, the operation of each party (A and B) would be well-localised in spacetime, and there is a clear sense in which these operations are taking place. They would enter a definite, yet cyclic causal order, with retrocausal influences taking place. This corresponds to the scenario assumed in section~\ref{TvsS}.

\end{enumerate}

\color{black}
In sum, among the various possible implementations of the QS (simulation, proper ICO with an indefinite spacetime, or proper ICO with a cyclic causal structure),\footnote{\textcolor{black}{As a note, both physical interpretations for the 'proper' realisations of the QS (namely that in terms of retrocausal influences and that in terms of metaphysically indeterminate orders) allow us to shed some light on the problematic meaning of the branched probabilities in the definition of the QS within the PMF. 
\newline
Let's denote $P^{A \preceq B}(X|Y)$ the probability distribution of obtaining outcome X conditioned on information Y in a given party within a process similar to the QS, but for which the control system would be in the definite state $\ket{\psi_c} = \ket{0_c}$. Similarly, we denote $P^{B \preceq A}(X|Y)$ the probability distribution of measurements' outcomes within a variant process for which $\ket{\psi_c} = \ket{1_c}$. In the QS, the generated probability distribution for measurements' outcomes is a probabilistic mixture of the branched ones: $P(X|Y) = \frac{1}{2}[P^{A \preceq B}(X|Y) + P^{B \preceq A}(X|Y)]$. This generated distribution is modelled within the PMF as the result of the superposed order between parties $A$ and $B$. As \citetbjps{adlam2023there} pointed out, the meaning of $P^{A \preceq B}(X|Y)$ and $P^{B \preceq A}(X|Y)$ is far from clear at first sight. The worry is that while a solution to the measurement problem assigns a specific meaning to the quantum probabilities, those relate to outcomes of measurements being the result of a collapse (objective or not) of the wavefunction, or at least belonging to decohered branches. However, in the present case, the probability distributions $P^{A \preceq B}(X|Y)$ and $P^{B \preceq A}(X|Y)$ are defined within distinct coherent branches that can interfere. To this day, no interpretation of probabilities in this context has been articulated. 
\newline
Now, let's emphasise that these probability distributions $P^{A \preceq B}(X|Y)$ and $P^{B \preceq A}(X|Y)$ are defined so that each is associated with a pair of events in a definite order and obeying relativistic causality. If one ontologically commits to the modelisation of the QS in terms of a definite order with retrocausal influences (i.e. the principle of relativistic causality is explicitly violated), as expressed in the framework of \citetbjps{vilasini2022embedding}, then these distributions do not refer to the objective reality. Instead, these distributions are merely theoretical devices, i.e. elements of a model that is empirically successful but ontologically wrong. 
\newline
If one chooses to stick to the definition of ICOs as defined in the PMF, and adopt a literal interpretation in terms of metaphysical indeterminacy of orders, then the superposition of orders in the QS means that it is metaphysically indeterminate which order obtains. The branched probabilities  $P^{A \preceq B}(X|Y)$ and $P^{B \preceq A}(X|Y)$ correspond to the single chances associated with a single run of a bipartite experiment similar to the QS, but in which a definite order obtains. Since, in the QS, the order is metaphysically indefinite, it is also indeterminate which single chance obtains in the world.}} only the 'proper' ICO instantiating an (output-impacting) cyclic relativistic causal structure would yield temporal nonlocality from a violation of a temporal Bell inequality.\footnote{\textcolor{black}{It can be shown that the evolution of the target system is non-unitary in any implementation of the QS, independently of the type of spatiotemporal embedding of the process. Although this paves the road for the possibility of temporally nonlocal correlations, this non-unitarity will not yield temporal nonlocality if there is no clear sense in which the involved events take place at all.}} The question is now whether one can assume, in a model-independent way, that we are in the presence of such a particular implementation of the QS. 

\textcolor{black}{The gravitational version of the QS (for which the control system is a mass configuration in a superposition of states, inducing a superposition of spacetime curvatures with opposite orders between the parties $A$ and $B$) is usually considered as a proper implementation of an ICO, if it were to be realised}: the metric itself (encoding the relativistic causal order between the operations) is either in a quantum superposition of states (as expressed in the PMF), or encodes a cyclic geometry (as might be expressed in the alternative framework of \citetbjps{vilasini2022embedding}) and so only two spatiotemporal events are involved (one for each operation) in either case. Yet, while the gravitational quantum switch is deemed 'in principle feasible', its realisation would be highly challenging on a technical level~\citepbjps{zych2019bell}. Moreover, this thought experiment relies on model-dependent claims, such as the theoretical existence of spacetime superpositions. Under what conditions these superpositions would be admissible in a theory of quantum gravity is not trivial~\citepbjps{paunkovic2020causal}. For our purposes, the gravitational QS would not need to ontologically correspond to a metaphysically indefinite spacetime, but rather to a definite, yet cyclic, spacetime structure. Similarly, one would need to have this scenario well described in a theory of quantum gravity.
%, and any description of the gravitational QS would ultimately be theory-laden, hence, model-dependent.
%
%ok - improve. Explain that if superpsitions of metrics, we can't have 4-events. Because this time, it is not that indefinite path of target so that it is indeterminate which ones is occuring. If 4 events with superposed metrics, we have four events that actually happens, but their order is indefinite.  if 4 spatiotemporal events, then we would have fours IT-parties in ICO, not just 2. 

\textcolor{black}{It might be impossible to obtain a ‘proper’ ICO, as described by the PMF in terms of an indefinite causal structure, without gravitational effects. If one considers \citetbjps{de2022quantum}’s reformulation of the definition of ICO in relativistic terms (which makes it applicable to the different types of QS implementations), it appears that an ICO boils down to a superposition of spatiotemporal descriptions, which can arise either via a superposition of metrics (as in the gravitational variant of the QS), or via a superposition of worldlines of some physical system (as in the flat spacetime versions of the QS, including the relativistic version of the QS proposed in \citepbjps{dkebski2022indefinite}). The latter case corresponds to a mere simulation of an ICO in fixed spacetime.}

\textcolor{black}{However, 'proper' ICOs instantiating a definite, yet cyclic structure, could in principle be obtained with or without gravitational effects. A retrocausal dynamics can indeed be conceived in a flat, fixed spacetime.} The theorem of \citetbjps{zych2019bell} has been adapted to an experimental setup in flat spacetime~\citepbjps{dkebski2022indefinite}, which further ensures that we do not need a gravitational setup to test for temporal nonlocality. 

\textcolor{black}{For the purpose of the discussion, let's assume that we accept that proper ICOs without gravitational effects are necessarily of the cyclic type.}
\textcolor{black}{When considering a version of \citetbjps{zych2019bell}'s experiment in flat-spacetime, a first worry is that there is still the possibility of ending up with a mere simulation of an ICO. It is arguably an open question whether there is still hope for the possibility of a model-independent distinction between a simulation and a proper ICO. 
According to \citetbjps{paunkovic2020causal}, it is experimentally possible, at least in principle, to discriminate between an implementation of the QS involving two or four spacetime events. However, an objection to this proposal has been voiced in \citepbjps{de2024event}, leaving the question open at this stage.}

\textcolor{black}{A second worry is that the argument inferring temporal nonlocality from a violation of a temporal Bell inequality would still rely on the modelisation of ICOs as instantiating an 'output-impacting' type of retrocausal model, which adds a further model-dependent stance. As discussed at the end of section~\ref{TvsS}, a natural variant of the very definition of temporal locality would alleviate this drawback.}

To conclude this section, the present protocol to test temporal locality is strictly speaking model-dependent. %add sum
This protocol has however the advantage that it would not require assuming a strong form of realism, contrary to the tests based on the Leggett-Garg and the temporal CHSH inequalities (see section~\ref{intro}). Similarly to the case of Bell theorem, this test would remain minimally restricting regarding (meta)physical assumptions other than temporal locality. %+ unicity of outcomes? and other axioms cavalcanti?"

\color{black}
\section{Distinction between Spatial and Temporal Nonlocality}\label{Dist}

\color{black}
As already emphasised by \citetbjps{adlam2018spooky}, a violation of a Bell inequality does not distinguish between a violation of locality between spacelike or timelike events. \textcolor{black}{The above discussion in section~\ref{dep} aligns with this observation, as it might be the case, with the experience of \citetbjps{zych2019bell}, that no model-independent version of the test is to exist to establish whether a violation of a Bell inequality between the spacelike-separated parties $C_1$ and $C_2$ is due to the presence of temporal nonlocality between the timelike-separated parties $A_i$ and $B_i$.} This raises interesting questions regarding the meaningfulness of a distinction between both kinds of nonlocality.
%an instance of spatial nonlocality between the spacelike-separated parties $C_1$ and $C_2$ with and without the presence 

The formal difference between spatial and temporal nonlocality boils down to the presence of non-classical correlations between spacelike-separated events for the former, and timelike-separated events for the latter. However, this different formal treatment between the two kinds of nonlocality does not imply that they are necessarily underpinned by distinct physical mechanisms.
%If only simulations of ICOs were to exist in nature, the
%then violations of the temporal Bell inequalities derived by \citepbjps{zych2019bell} would amount solely to standard Bell nonlocality between spacelike-separated events.  

If proper ICOs can be implemented in nature, then they would refer to something objective in the world, not captured by the mathematical apparatus of standard QM. The physicality of proper ICOs would therefore point towards the need for a formalism that treats time and space in a more unified manner. This unified treatment would also support a possibly (new and) similar physical underpinning for both spatial and temporal nonlocality (if the latter is applicable in the first place). 
%
%of which the physical interpretation would explain in a unified way the instances of nonlocal correlations. 

In the context of \citepbjps{zych2019bell}'s setup, no matter the type of implementation for the ICOs, spatial nonlocality between parties $C_1$ and $C_2$ would be a consequence of the presence of these ICOs. In case we assume proper ICOs, spatial (and possibly temporal nonlocality if applicable) would be provided a (unified) explanation through retrocausal influences (e.g., retrocausality has already been used to account for standard nonlocality~\citepbjps{sep-qm-retrocausality}), or as originating from some form of metaphysical indeterminacy of spatiotemporal relations.\footnote{Other realist readings of proper ICOs could possibly lead to further alternative conclusions. Aside from the above-mentioned reading of ICOs, a representationalist approach could also commit to some kind of global holism affecting the dynamics or the ontology of the world across time and space.
\newline
Alternatively, one could also adopt a non-representational approach to indefinite causal orders, and take the probability distributions within the QS as not determined by objective properties of the systems and the (order of) quantum events, but as characterising our relation to these aspects. While a representationalist interpretation towards ICO will locate temporal nonlocality in objective reality, a non-representationlist approach will locate it at the level of our relation to the external world (without it being a mind-dependent feature for all that).} This discussion highlights the fact that focusing on ICOs and their physical meaning could support currently underexplored ways to understand not only temporal nonlocality (if applicable), but Bell nonlocality as well.

%then either temporal nonlocality can occur as the result of non-signalling retrocausal influences, or it is plainly inapplicable as a concept as the result of metaphysical indeterminacy of events('orders). In both cases,
\color{black} 

\section{Conclusion}\label{ccl}

%Conclusions may be used to restate your hypothesis or research question, restate your major findings, explain the relevance and the added value of your work, highlight any limitations of your study, describe future directions for research and recommendations. 

This work investigated to what extent ICOs can be used as a key ingredient to test the principle of temporal locality, as defined in \citepbjps{adlam2018spooky}. \textcolor{black}{It was shown that a violation of the temporal Bell inequalities derived from the protocol proposed in \citepbjps{zych2019bell} could imply temporal nonlocality modulo model-dependent assumptions.
In particular, the presence of temporally nonlocal correlations would be explained by a specific implementation of ICOs involving output-impacting retrocausal influences. Other implementations of ICOs would have different implications for the existence and applicability of temporal nonlocality.} On the other hand, unlike other tests based on the Leggett-Garg and the temporal CHSH inequalities, this protocol would be free from any commitment to a strong form of realism, allowing the test to effectively target a well-defined principle of temporal locality. 
It was also argued that this new emphasis on ICOs can not only help explore the possibility of temporal nonlocality, but can also support new interpretational routes for standard quantum nonlocality, based on a more unified treatment of the spatial and temporal dimensions. 
%Future research could explore the (im)possibility to render the present test fully model-independent, and identify a suitable notion of (spatio)temporal entanglement necessary for Adlam's temporal nonlocality, in the hope it would further shed light on the nature of the distinction between spatial and temporal nonlocality. The work of \citepbjps{bera2019quantifying}, which studied how certain types of evolution's superpositions could prove necessary to the violation of the temporal Bell inequalities of \citepbjps{zych2019bell}, might be an interesting step in that direction. 

%Prospects to render the inference of temporal nonlocality fully model-independent might require proving that ICOs in fixed, flat spacetime would necessarily correspond to a causal structure violating local causality, and their faithful implementation should be guaranteed model-independently. 

\section*{Acknowledgments}

I am grateful to Vincent Lam and Cyril Branciard for discussions at the early stages of this work. Many thanks to Marco Fellous Asiani and Hippolyte Dourdent for helpful technical clarifications regarding the quantum switch. This work also benefited from valuable exchanges with the audience at the following events: the philosophy of physics seminars from the Warsaw University of Technology, the Munich Center for Mathematical Philosophy, the University of Groningen and the University of Geneva, the 21st European Conference on Foundations of Physics (Bristol), the 4th Lisbon International Conference on Philosophy of Science and the 9th biennial meeting of the European Philosophy of Science Association (EPSA23). In particular, I would like to thank Emily Adlam for very helpful feedback on a previous version of this work. Finally, I am grateful to the anonymous referees that provided me with precious suggestions for improvements.

This project has received funding from the Formal Epistemology Project funded by the Czech Academy of Science, and the European Union’s Horizon 2022 research and innovation programme under the Marie Skłodowska-Curie grant agreement No [101109486].

  \begin{figure}[H]
    \includegraphics[width=0.5\textwidth]{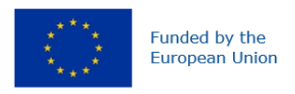}
    \end{figure}

\begin{FlushRight}
Munich Center for Mathematical Philosophy

Ludwig-Maximilians-Universität München

München, Germany
\end{FlushRight}

\bibliographystyle{bjps}
\bibliography{sn-bibliography}

\end{document}